\documentclass[fleqn,usenatbib]{mnras}

\usepackage{newtxtext,newtxmath}
\usepackage[T1]{fontenc}

\DeclareRobustCommand{\VAN}[3]{#2}
\let\VANthebibliography\thebibliography
\def\thebibliography{\DeclareRobustCommand{\VAN}[3]{##3}\VANthebibliography}

\usepackage{graphicx}
\usepackage{amsmath}
\usepackage{xcolor}
\usepackage{multirow}
\usepackage{subcaption}

\title[Interacting Dark Sectors with DESI DR2]{\boldmath Interacting Dark Sectors in light of DESI DR2}

\author[R. Shah, P. Mukherjee, S. Pal]{
Rahul Shah,$^{1}$\thanks{E-mail: rahul.shah.13.97@gmail.com}
Purba Mukherjee,$^{1,2}$\thanks{E-mail: purba16@gmail.com}
and Supratik Pal$^{1}$\thanks{E-mail: supratik@isical.ac.in}
\\
$^{1}$Physics and Applied Mathematics Unit, Indian Statistical Institute, 203 B.T. Road, Kolkata 700 108, India\\
$^{2}$Centre for Theoretical Physics, Jamia Millia Islamia, New Delhi 110025, India
}

\begin{document}
\label{firstpage}
\pagerange{\pageref{firstpage}--\pageref{lastpage}}
\maketitle

\begin{abstract}
Possible interaction between dark energy and dark matter has previously shown promise in alleviating the clustering tension, without exacerbating the Hubble tension, when Baryon Acoustic Oscillations (BAO) data from the Sloan Digital Sky Survey (SDSS) DR16 is combined with Cosmic Microwave Background (CMB) and Type-Ia Supernovae (SNIa) data sets. With the recent Dark Energy Spectroscopic Instrument (DESI) BAO DR2, there is now a compelling need to re-evaluate this scenario. We combine DESI DR2 with Planck 2018 and Pantheon+ SNIa data sets to constrain interacting dark matter dark energy models, accounting for interaction effects in both the background and perturbation sectors. Our results exhibit similar trends to those observed with SDSS, albeit with improved precision, reinforcing the consistency between the two BAO data sets. In addition to offering a resolution to the $S_8$ tension, in the phantom-limit, the dark energy equation of state exhibits an early-phantom behaviour, aligning with DESI DR2 findings, before transitioning to $w\sim-1$ at lower redshifts, regardless of the DE parametrization. However, the statistical significance of excluding $w=-1$ is reduced compared to their non-interacting counterparts.
\end{abstract}



\section{Introduction}

Cosmology today is an observationally driven field that has made tremendous progress in recent years, thanks to expansive and novel data from cutting-edge instruments and observatories. In particular, the Dark Energy Spectroscopic Instrument (DESI)\footnote{\url{https://www.desi.lbl.gov/}} has reignited excitement by hinting at a dynamically evolving dark energy (DE) equation of state (EoS), which challenges the standard cosmological constant ($\Lambda$) paradigm \citep{DESI:2024mwx}. This has led to a surge of studies investigating DESI's apparent non-$\Lambda$ preference \citep{Giare:2024gpk,Dinda:2024ktd,Pourojaghi:2024bxa,Colgain:2024mtg,Bansal:2025ipo,Sousa-Neto:2025gpj,Luongo:2024fww,Cortes:2024lgw,Colgain:2024xqj,Chan-GyungPark:2024mlx,DESI:2024aqx,DESI:2024kob,Roy:2024kni,Gialamas:2024lyw,Notari:2024rti,Chakraborty:2025syu,Chakraborty:2024xas,Chan-GyungPark:2025cri,Khoury:2025txd,Mukherjee:2024ryz,Mukherjee:2025fkf}, with the recent data release 2 (DR2) \citep{DESI:2025zgx} continuing this trend.

On top of the excitement surrounding evolving dark energy, DESI DR2 also provides the most precise and up-to-date baryon acoustic oscillations (BAO) data set \citep{DESI:2025zgx}. This warrants a re-evaluation of key previous results, especially those based on the SDSS BAO data, by the authors of the present paper \citep{Shah:2024rme}. The tensions afflicting the $\Lambda$CDM model are now well-known, with the Hubble tension and clustering tension inspiring hundreds of studies on alternative physical models, experimental systematics, and novel analysis pipelines \citep{DiValentino:2021izs,DiValentino:2020vvd,Abdalla:2022yfr}. Among these, interactions in the dark sector, i.e., between dark matter and dark energy, have been extensively explored in the literature (see \citet{Pan:2023mie} and references therein). Certain interaction set-ups have shown promise in alleviating specific cosmological tensions (see \citet{Shah:2024rme,Li:2024qso,Giare:2024ytc,Giare:2024smz,Zhai:2025hfi,Bhattacharyya:2018fwb,Sinha:2021tnr} and references therein).


\section{Interacting Set-Up and Methodology}

We consider the interacting dark matter dark energy (iDMDE) set-up outlined in Sec. 2 of \citet{Shah:2024rme}, which is compatible with dynamical dark energy (DDE) scenarios. This makes it well-suited for re-examination in light of DESI DR2, given the mild preference for an EoS deviating from $w=-1$. Here, we replace the SDSS DR16 BAO data used in \citet{Shah:2024rme} with DESI DR2. We constrain the models using a combination of Planck 2018 TTTEEE + low-$\ell$ + low-E + lensing data \citep{Planck:2019nip,Planck:2018vyg,Planck:2018lbu}, DESI DR2 BAO distance measurements (as listed in Table IV of \citet{DESI:2025zgx}), and Pantheon+ SNIa compilation \citep{Scolnic:2021amr}. We employ a modified version of \texttt{CLASS} \citep{Lesgourgues:2011re,Blas:2011rf} (derived from \citet{Hoerning:2023hks,Lucca:2020zjb}) and \texttt{MontePython} \citep{Brinckmann:2018cvx,Audren:2012wb}, with the DESI DR2 likelihood internally developed and cross-checked for consistency with results obtained from the DESI DR2 likelihood in \texttt{Cobaya} \citep{Torrado:2020dgo}.

\begin{table*}
    \resizebox{1.0\textwidth}{!}{\renewcommand{\arraystretch}{1.2} \setlength{\tabcolsep}{25 pt}
    \begin{tabular}{c c c c c}
        \hline\hline
        \textbf{Parameters} & \textbf{CPL} & \textbf{i-CPL} & \textbf{JBP} & \textbf{i-JBP} \\ \hline
        {\boldmath${\Omega_b}{h^2}$} & $0.02244\pm 0.00014        $ & $0.02246\pm 0.00013        $ & $0.02250\pm 0.00014        $ & $0.02253\pm 0.00013        $\\
        {\boldmath${\Omega_c}{h^2}$} & $0.11905\pm 0.00087        $ & $0.1558^{+0.0087}_{-0.0071}$ & $0.1184\pm 0.0012          $ & $0.151^{+0.012}_{-0.0097}  $\\
        {\boldmath$100{\theta_s}$} & $1.04199\pm 0.00029        $ & $1.04202\pm 0.00028        $ & $1.04204\pm 0.00029        $ & $1.04210\pm 0.00027        $\\
        {\boldmath${\ln{\left({10^{10}A_s}\right)}}$} & $3.045\pm 0.014            $ & $3.048\pm 0.015            $ & $3.051\pm 0.016            $ & $3.053\pm 0.016            $\\
        {\boldmath$n_s$          } & $0.9677\pm 0.0034          $ & $0.9684\pm 0.0036          $ & $0.9695\pm 0.0039          $ & $0.9702\pm 0.0035          $\\
        {\boldmath${\tau}$       } & $0.0558\pm 0.0071          $ & $0.0570^{+0.0071}_{-0.0080}$ & $0.0585\pm 0.0080          $ & $0.0601^{+0.0081}_{-0.0091}$\\
        {\boldmath$Q$            } &              -               & $0.390^{+0.10}_{-0.084}    $ &               -              & $0.34^{+0.13}_{-0.11}      $\\
        {\boldmath$w_0$          } & $-0.837\pm 0.056           $ & $> -1.04                   $ & $-0.807\pm 0.085           $ & $> -1.06                   $\\
        {\boldmath$w_a$          } & $-0.59^{+0.23}_{-0.21}     $ & $-0.33\pm 0.15             $ & $-1.09\pm 0.53             $ & $-0.33\pm 0.30             $\\
        \hline
        {\boldmath$H_0$          } & $67.57\pm 0.59             $ & $67.92\pm 0.52             $ & $67.57\pm 0.70             $ & $68.07\pm 0.53             $\\
        {\boldmath$\Omega_{m0}$  } & $0.3114\pm 0.0057          $ & $0.388^{+0.021}_{-0.018}   $ & $0.3102\pm 0.0072          $ & $0.376^{+0.027}_{-0.023}   $\\
        {\boldmath$\sigma_{8,0}$ } & $0.8098\pm 0.0090          $ & $0.661^{+0.024}_{-0.031}   $ & $0.806\pm 0.011            $ & $0.673^{+0.034}_{-0.044}   $\\
        {\boldmath$S_8$          } & $0.8250\pm 0.0098          $ & $0.751^{+0.014}_{-0.015}   $ & $0.819\pm 0.012            $ & $0.751^{+0.018}_{-0.021}   $\\
        \hline 
        {\boldmath $\chi^2_{min}$ } & 4199 & 4201 & 4202 & 4204 \\
        {\boldmath $-\ln{\cal L}_\mathrm{min}$ } & 2099.70 & 2100.32 & 2100.84 & 2102.04 \\
        \hline
        \hline
    \end{tabular}
    }
\caption{The mean and 1$\sigma$ constraints obtained for interacting (phantom regime) and non-interacting models considered in this work, using combined Planck 2018 + DESI DR2 BAO + Pantheon+ data sets.}
\label{tab:desitable}
\end{table*}

Furthermore, when considering perturbations in both dark sectors to investigate the impact of evolving DE on clustering, it is well-known that the perturbation equations contain a term (the ``doom factor" \citep{Gavela:2009cy}) in the denominator that diverges at the phantom line ($w = -1$) \citep{Gavela:2010tm}. To extract meaningful information, it is therefore customary to perform the analysis separately for the phantom and non-phantom regions (as described in \citet{Bhattacharyya:2018fwb, Shah:2024rme}). In the non-interacting case, this crossing is handled smoothly using the parametrized post-Friedmann (PPF) formalism \citep{Fang:2008sn} (as also done by the DESI collaboration). However, in interacting models, the energy exchange between dark matter and dark energy introduces additional terms in the perturbation equations. To avoid degenerate effects from combining PPF with the interaction term, we divide the parameter space into non-phantom $w(z)>-1$ and phantom $w(z)<-1$ regimes. This also allows for distinct prior ranges on the DE EoS parameter, $w_0$, while preserving the exact perturbation equations. In our analysis, we impose only this phantom bound, deliberately avoiding any further constraints on the interaction parameter or related quantities. This allows us to directly test for potential theoretical instabilities using current data, rather than pre-emptively excluding them employing hard bounds (as done by many past authors), and thus draw more general and data-driven conclusions.
Additionally, we fix the sound speed of dark energy perturbations to $c_s^2 = 1$.

Given the $S_8$ values reported in \citet{Li:2023azi,DES:2021wwk,Kilo-DegreeSurvey:2023gfr}, the results from SDSS BAO \citep{Shah:2024rme} indicate that a phantom EoS helps alleviate the $S_8$ tension without worsening the $H_0$ tension. Whereas, a non-phantom EoS tends to exacerbate the $S_8$ tension, and slightly worsen the $H_0$ tension. We find similar conclusions with DESI DR2 regarding the status of $H_0-S_8$ tensions. Because of the previous success of the phantom regime and its continued validity with the latest data sets, we primarily focus on phantom results. For the non-phantom set-up, we refer the reader to Sec. \ref{non-phantom}, where we outline its key characteristics.

As representative DDE models, we consider two widely accepted parametrizations of the DE EoS: the Chevallier-Polarski-Linder (CPL) [$w(a) = w_0 + w_a (1 - a)$] \citep{cpl1,eos1} and the Jassal-Bagla-Padmanabhan (JBP) [$w(a) = w_0 + w_a a (1 - a)$] \citep{jbp}. The CPL and JBP parametrizations are most widely utilized in dynamical DE studies, providing greater flexibility compared to simpler models such as $w$CDM, for analysing the evolution of the DE EoS. Moreover, we choose these two particular parametrizations, due to their demonstrated success in alleviating the $S_8$ tension with SDSS BAO in an interacting set-up \citep{Shah:2024rme}.

To ensure the consistency of our \texttt{MontePython} likelihood for DESI DR2 with the \texttt{Cobaya} likelihood used by the DESI collaboration, we have first validated the non-interacting cases for the same parametrizations. The constraints obtained for the non-interacting cases (particularly CPL) are found to be fully consistent with those reported by the DESI collaboration \citep{DESI:2025zgx, DESI:2025fii}, confirming the reliability of our pipeline for further exploration of the interacting sectors using \texttt{CLASS} + \texttt{MontePython}.

\vspace{-15pt}


\section{Results and Discussion}

\begin{figure*}
\begin{center}
    \begin{subfigure}{.48\textwidth}
        \includegraphics[width=\textwidth]{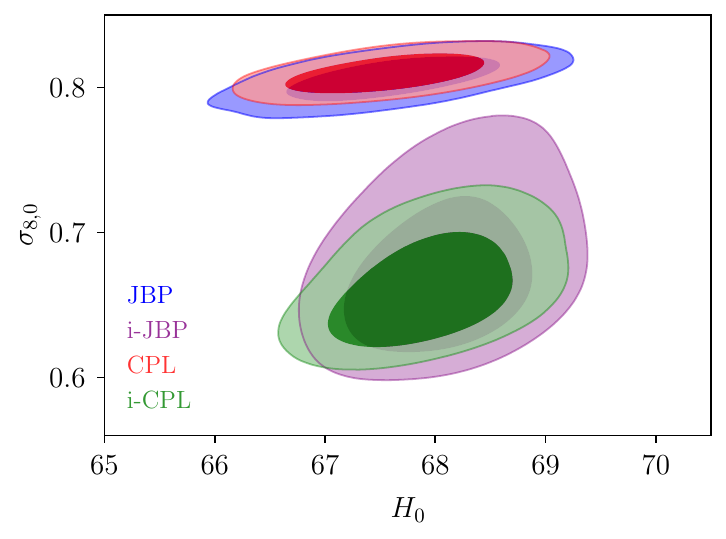}
    \end{subfigure}
    \begin{subfigure}{.48\textwidth}
        \includegraphics[width=\textwidth]{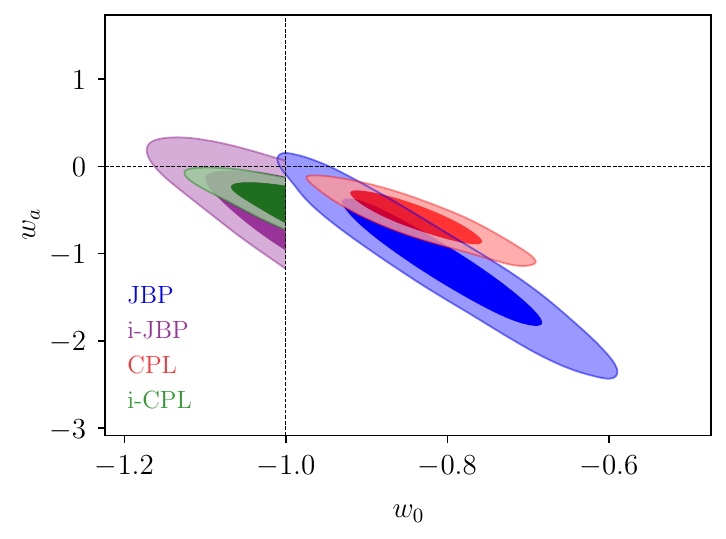}
    \end{subfigure}
\end{center}
    \vspace{-10pt}
    \caption{Constraints on and correlations between $H_0$ and $\sigma_{8,0}$, as well as $w_0$ and $w_a$, are shown for the models considered in this work. The non-interacting cases smoothly incorporate $w = -1$ crossing using the PPF formalism. In contrast, the interacting cases divide the parameter space into phantom and non-phantom regions - only the phantom case is shown here.}
    \label{fig:2Dplots}
\end{figure*}

Table \ref{tab:desitable} presents the constraints for both parametrizations, along with their interacting counterparts (denoted by the prefix `i-'). We present the full contour plots in Fig. \ref{fig:triangle}, which were generated using \texttt{GetDist} \citep{Lewis:2019xzd}. We make the following observations:

\begin{itemize}
    \item \textbf{Consistency between SDSS and DESI DR2:} Comparing Table \ref{tab:desitable} with Table III of \citet{Shah:2024rme}, we observe minor shifts in $H_0$, $\Omega_{m0}$, $\sigma_{8,0}$, and $S_8$. However, these shifts are statistically insignificant, with all constraints remaining consistent to within $1\sigma$ between the SDSS and DESI DR2 cases, along with a slight overall increase in precision. Furthermore, a comparison between Fig. \ref{fig:triangle} and Fig. 5 of \citet{Shah:2024rme} reveals identical trends and correlations.
    \vspace{0.5\baselineskip}
    \item \textbf{The tensions:} The Hubble tension is neither alleviated nor worsened in the case with DESI compared to SDSS, although a slight increase in the mean value of $H_0$ is observed (supposedly driven by DESI). The clustering tension is alleviated in a similar manner to that described in \citet{Shah:2024rme} for DESI DR2, with slightly lower mean values for $S_8$ compared to SDSS, while maintaining the same level of precision within the $1\sigma$ confidence level. The presence of interaction reduces the $H_0$–$\sigma_{8,0}$ correlation, as shown in the left panel of Fig. \ref{fig:2Dplots}, a trend that remains consistent across both BAO data sets.
    \vspace{0.5\baselineskip}
    \item \textbf{Nature of dark energy:} The constraints on the EoS parameters ($w_0$ and $w_a$) from SDSS and DESI DR2 are largely consistent for the interacting cases. As shown in Fig. \ref{fig:2Dplots}, both interacting and non-interacting models favour a deviation from $w=-1$. However, in the phantom limit, the i-CPL and i-JBP models prefer a less strong phantom-like evolution at early times. For CPL, $w_0 = 0.837 \pm 0.056$ and $w_a = -0.59^{+0.23}_{-0.21}$, leading to $w \approx -1.425$ at high redshifts (see \citet{DESI:2025zgx, DESI:2025fii}). In contrast, i-CPL shifts the $w_0 > -1.04$ and $w_a = -0.33 \pm 0.15$, making the early universe EoS less phantom ($w \approx -1.347$). For JBP, $w_0 = -0.807\pm 0.085$ and $w_a = -1.09 \pm 0.53$, whereas i-JBP gives $w_0 > -1.06$, $w_a = -0.33\pm 0.3$. This suggests that the iDMDE models suppress extreme phantom evolution in the early universe while still allowing a transition toward $w\sim-1$ at lower redshifts.
    \vspace{0.5\baselineskip}
    \item \textbf{Nature of interaction:} The presence of a positive, non-zero $Q$ in both interacting models means that dark energy is decaying into dark matter. The interaction helps to gradually suppress deviations from $w=-1$ at early times compared to non-interacting CPL/JBP models.
    \vspace{0.5\baselineskip}
    \item \textbf{Value of $\Omega_{m0}$:} It should be noted that the interacting cases with $Q>0$ tend to prefer a higher value of $\Omega_{m0}$ which is in $\sim 4.05\sigma$ tension with the non-interacting CPL, $\sim 2.73\sigma$ tension with the non-interacting JBP, as well as in $3.78 \sigma$ and $2.54 \sigma$ tension with Planck $\Lambda$CDM case, respectively. A closer examination of the observed correlations reveals that a positive $Q$ implies an injection of energy into the matter sector, thereby increasing the value of $\Omega_{m0}$. Consequently an increase in the value of $\Omega_{m0}$ is complemented with a decrease in $\sigma_{8,0}$ and $S_8$.
\end{itemize}

When interactions are introduced, different results for $\sigma_{8,0}$ (and to a lesser extent, $H_0$) emerge in the two regimes. We find that the phantom interacting case helps alleviate the $S_8$ tension when using the latest DESI DR2 BAO data, with a shift in the mean value towards a lower $S_8$ (a direction favoured towards addressing the tension). Sec. \ref{non-phantom} shows that $S_8$ assumes a higher mean value in the non-phantom case (consistent with our previous findings using SDSS BAO \citep{Shah:2024rme}), but it can accommodate lower $S_8$ values due to reduced precision in the inferred constraint. Thus, despite the preference for a present-day non-phantom EoS in the non-interacting case (with DE perturbations handled using the PPF formalism) across both BAO data sets, the phantom regime appears to yield cosmological parameter estimates, particularly for $H_0$ and $S_8$, that better align with observational constraints when both interactions and dark energy perturbations are considered.

\begin{figure}
    \centering
    \includegraphics[width=0.48\textwidth]{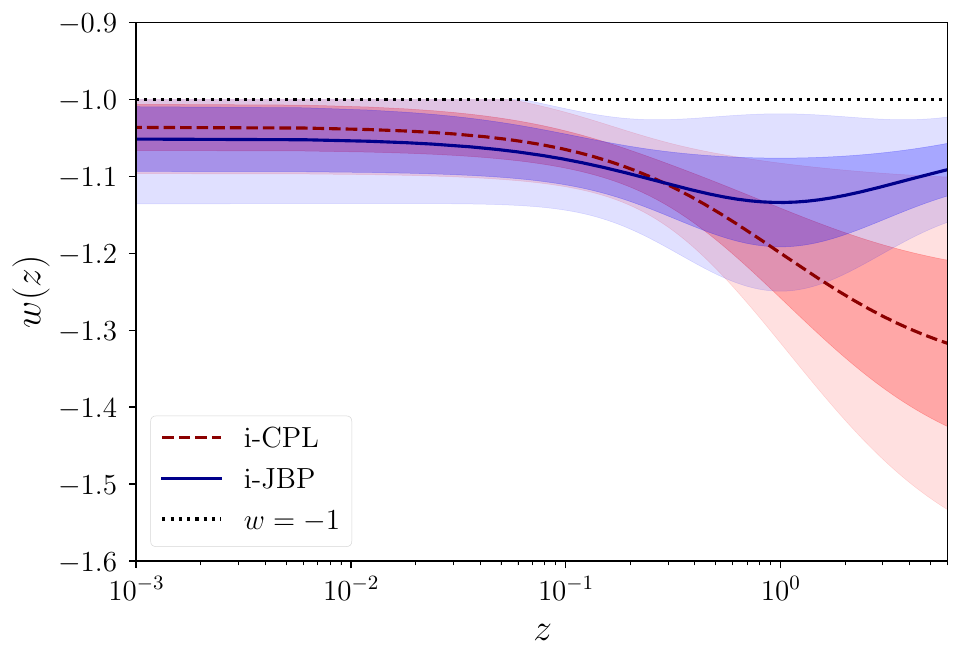}
    \caption{Plots for the evolution of DE EoS for interacting CPL and JBP models. The best-fitting line is shown along with the 1$\sigma$ and 2$\sigma$ confidence intervals in the shaded regions.}
    \vspace{-10pt}
    \label{fig:EoS}
\end{figure}

\begin{figure*}
    \centering
    \includegraphics[width=\textwidth]{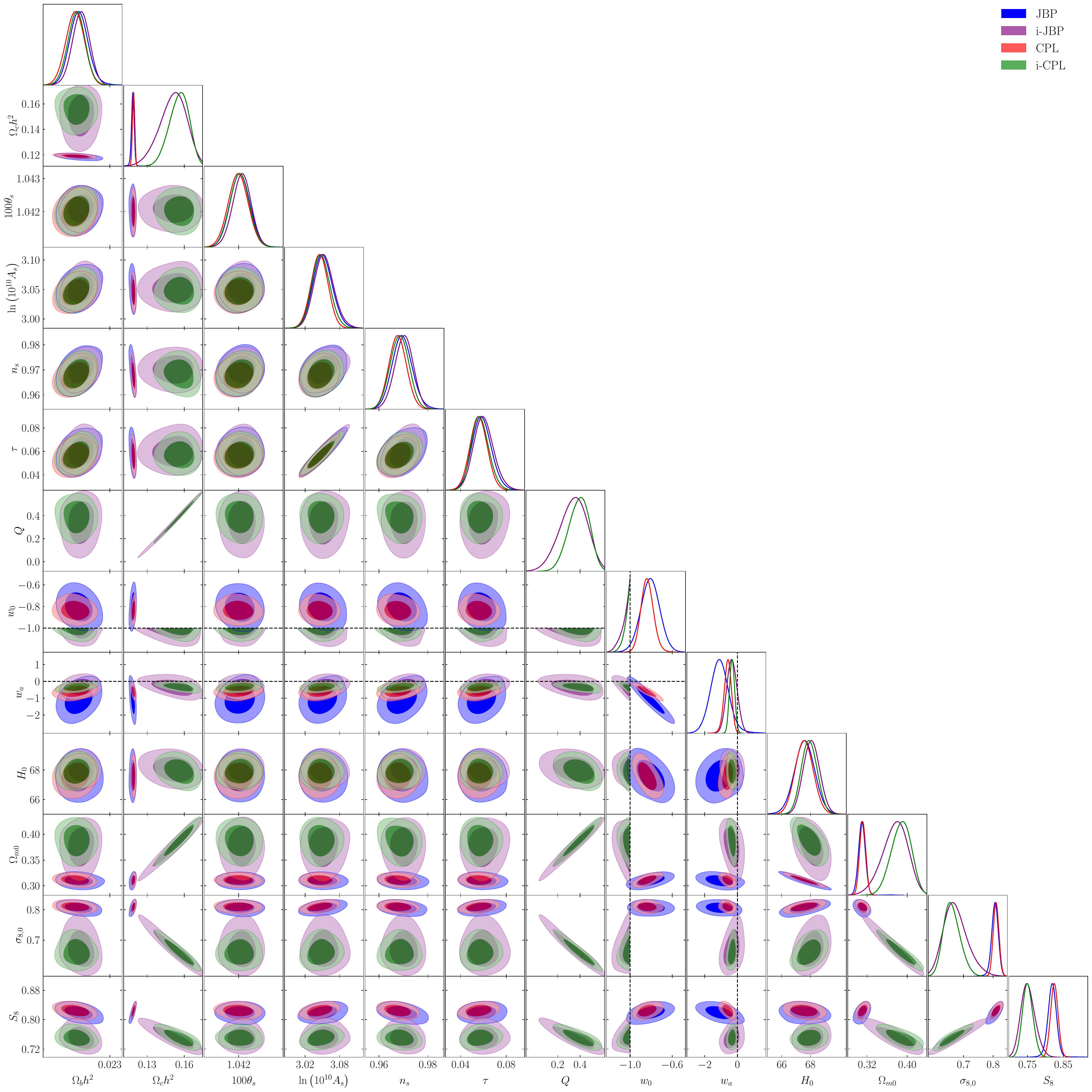}
    \caption{Comparison of constraints obtained for interacting (phantom regime) and non-interacting models considered in this work, using combined Planck 2018 + DESI DR2 BAO + Pantheon+ data sets.}
    \label{fig:triangle}
\end{figure*}

Notably, the surprise comes with the value of $w_a$, which largely governs the steep evolution of the EoS with DESI, making it tend to deviate from $w = -1$ case to a considerable extent at certain redshift regions of its evolution for the non-interacting CPL and JBP models. However, we find that the presence of interactions in the dark sector moderates this steep evolution, making the $w=-1$ deviation less pronounced compared to the non-interacting cases (see Fig. \ref{fig:EoS}).

At this stage, we interpret this moderation as a combined effect of interactions in the DM-DE sector and dark energy perturbations. Determining which of the two plays a more significant role is a matter for future investigation. However, since the inclusion of perturbations in the non-interacting set-up still indicates a deviation from the $w = -1$ EoS, we anticipate that the primary factor behind reducing the steep evolution is the interaction between the dark sectors. 

This leads to an interesting consideration: if DESI's preference for a steeply evolving EoS is indeed significant, it is worth exploring why this preference weakens in the presence of dark sector interactions - a paradigm that remains observationally viable and has shown promise in addressing cosmological tensions. 

Indeed, the interaction parameter $Q$ plays a significant role here, as it is seen that any large deviation from $w_a=0$ in the non-interacting case (as reported by the DESI collaboration) is now being compensated somewhat by the presence of a non-vanishing $Q$. Notably, the introduction of the interaction term does not worsen the fit to the data, as indicated by the $\chi^2$ values in Table \ref{tab:desitable}. Model selection criteria such as Akaike Information Criterion (AIC) \citep{1100705}, Bayesian Information Criterion (BIC) \citep{10.1214/aos/1176344136}, or the Bayes factor \citep{Kass01061995} naturally penalize additional parameters, like the interaction term $Q$, unless they yield significant reduction in $\chi^2$. We show a Bayesian evidence \citep{Trotta:2008qt} analysis of the models considered in this work in Appendix \ref{appendix}. However, our analysis indicates that dynamical interacting scenarios remain physically motivated, particularly due to their potential to ease the clustering tension. Furthermore, recent DESI DR2 results hint at DDE, reinforcing the relevance of probing possible interactions in the dark sector and their interplay with a time-varying dark energy component.

\begin{figure*}[!h] 
    \centering
    \includegraphics[width=0.98\textwidth]{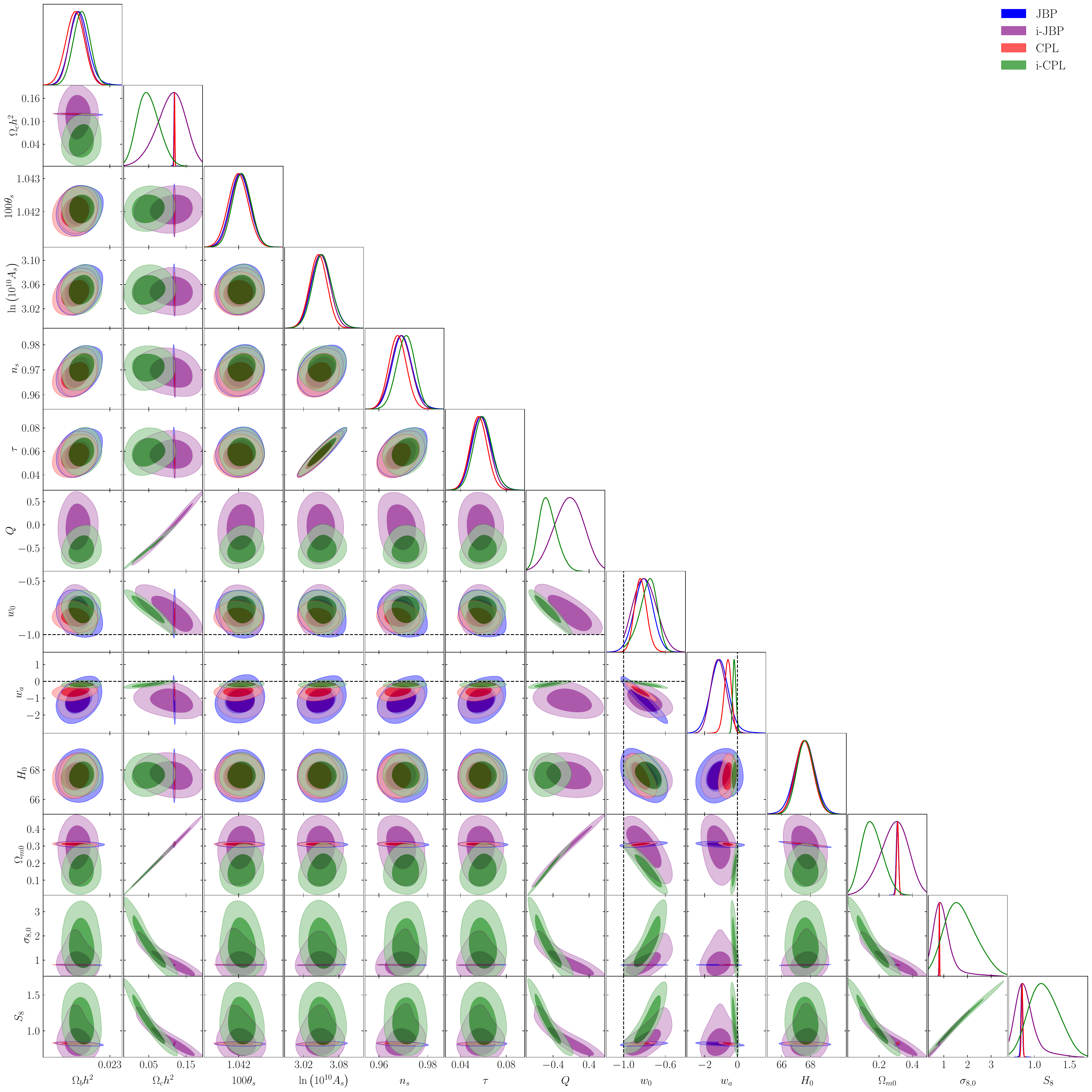}
    \caption{Comparison of constraints obtained for interacting (non-phantom regime) and non-interacting models considered in this work, using combined Planck 2018 + DESI DR2 BAO + Pantheon+ data sets.}
    \label{fig:triangle_quint}
\end{figure*}

Although these conclusions are based on a specific (albeit widely accepted) form of the interaction parameter, it is worthwhile to explore alternative interaction models and examine their impact on cosmological tensions and EoS parameters to draw more general conclusions about iDMDE sectors using DESI DR2 along with other data sets. Examples of such interactions can be found in \citet{Wang:2024vmw} and references therein.


\section{Non-Phantom Interacting Case}\label{non-phantom}

Here we show the constraints in the non-phantom regime for the interacting and non-interacting CPL and JBP cases for representative purposes. The full contour plots are in Fig. \ref{fig:triangle_quint} and the constraints are in Table \ref{tab:desitablequint}. We note that there is a peak in the posterior of $w_0$, unlike the phantom case. There is a negative correlation between $w_a$ and $Q$ in the case of i-CPL, indicating that deviations in $w_a$ from $0$ can be compensated by negative values of $Q$. This behaviour arises as a consequence of restricting our analysis to the non-phantom region, whereas the non-interacting CPL parametrization exhibits an early-phantom-like trend. In contrast JBP shows no such correlation and the posterior of $Q$ is centred near zero. However, $S_8$ takes a large value in both i-CPL and i-JBP with almost an order of magnitude worse precision than in the corresponding phantom case. Although this reduces the Gaussian tension on account of larger error bars, the direction of shift in the mean value of $S_8$ raises doubts on the admissibility of this scenario as a coherent resolution to cosmological tensions.

\begin{table}
    \resizebox{0.5\textwidth}{!}{\renewcommand{\arraystretch}{1.4} \setlength{\tabcolsep}{15 pt}
    \begin{tabular}{c c c}
        \hline\hline
        \textbf{Parameters} & \textbf{i-CPL Non-phantom} & \textbf{i-JBP Non-phantom} \\ \hline
        {\boldmath${\Omega_b}{h^2}$} & $0.02254\pm 0.00013        $ & $0.02248\pm 0.00013        $\\
        {\boldmath${\Omega_c}{h^2}$} & $0.049^{+0.026}_{-0.032}   $ & $0.110^{+0.042}_{-0.034}   $\\
        {\boldmath$100{\theta_s}$} & $1.04209\pm 0.00027        $ & $1.04207\pm 0.00028        $\\
        {\boldmath${\ln{\left({10^{10}A_s}\right)}}$} & $3.051\pm 0.015            $ & $3.048\pm 0.015            $\\
        {\boldmath$n_s$          } & $0.9709\pm 0.0035          $ & $0.9690\pm 0.0038          $\\
        {\boldmath${\tau}$       } & $0.0595\pm 0.0077          $ & $0.0574\pm 0.0077          $\\
        {\boldmath$Q$            } & $-0.52^{+0.17}_{-0.22}     $ & $-0.05\pm 0.30             $\\
        {\boldmath$w_0$          } & $-0.764^{+0.093}_{-0.074}  $ & $-0.787^{+0.093}_{-0.12}   $\\
        {\boldmath$w_a$          } & $-0.184^{+0.090}_{-0.11}   $ & $-1.15\pm 0.43             $\\
        \hline
        {\boldmath$H_0$          } & $67.66\pm 0.57             $ & $67.63\pm 0.60             $\\
        {\boldmath$\Omega_{m0}$  } & $0.158^{+0.056}_{-0.070}   $ & $0.291^{+0.093}_{-0.076}   $\\
        {\boldmath$\sigma_{8,0}$ } & $1.71^{+0.58}_{-0.77}      $ & $0.95^{+0.24}_{-0.44}      $\\
        {\boldmath$S_8$          } & $1.14^{+0.20}_{-0.23}      $ & $0.871^{+0.095}_{-0.17}    $\\
        \hline 
        {\boldmath $\chi^2_{min}$ } & 4206 & 4203 \\
        {\boldmath $-\ln{\cal L}_\mathrm{min}$ } & 2102.98 & 2101.29 \\
        \hline
        \hline
    \end{tabular}
    }
\caption{The mean and 1$\sigma$ constraints obtained for interacting CPL model in the non-phantom regime, using combined Planck 2018 + DESI DR2 BAO + Pantheon+ data sets.}
\label{tab:desitablequint}
\end{table}


\section{Conclusions}

In conclusion, in this study we revisited the iDMDE sector scenario in light of the latest DESI DR2 BAO measurements. We found that while DESI’s non-interacting dark energy EoS significantly deviates from $\Lambda$CDM, the inclusion of interactions in the dark sector moderates this deviation by reducing the preference for a steeply evolving EoS. The interaction plays a key role by reducing the $H_0$–$\sigma_{8,0}$ correlation and easing the clustering tension. Since the phantom regime induces a shift in $S_8$ towards lower values, thereby helping to address the clustering tension without worsening the Hubble tension, it remains the more interesting case; the non-phantom case on the other hand caused a shift in the other direction with lesser precision in $S_8$.

Our results underscore the importance of consistently incorporating perturbations in phantom models, and we suggest that exploring alternative interaction forms could provide further insights into dark sector dynamics, especially in light of novel data sets such as DESI DR2.

\textit{Note:} Although the DES-Y3 \citep{DES:2025xii} and KiDS-Legacy \citep{Wright:2025xka} collaborations have recently reported higher values for $S_8$, their new methodology awaits further reassessment through a systematic analysis by the community, considering all potential sources of uncertainty. In this article, we rely on the commonly accepted values of $S_8$ \citep{Li:2023azi,DES:2021wwk,Kilo-DegreeSurvey:2023gfr,Hamana:2019etx}.


\section*{Acknowledgements}

We thank Debarun Paul, Raj Kumar Das, Amlan Chakraborty and Amrita Mukherjee for useful discussions. We also thank the anonymous reviewer for their invaluable suggestions towards the improvement of the manuscript. RS acknowledges financial support from the Indian Statistical Institute (ISI), Kolkata as a Senior Research Fellow. PM acknowledges financial support from the Anusandhan National Research Foundation (ANRF), Govt. of India under the National Post-Doctoral Fellowship (N-PDF File no. PDF/2023/001986). SP thanks the ANRF, Govt. of India for partial support through Project No. CRG/2023/003984. We acknowledge the use of the Pegasus cluster of the high performance computing (HPC) facility at Inter-University Centre for Astronomy and Astrophysics (IUCAA), Pune, India.


\section*{Data Availability}

The data sets used in this work are all publicly available. The modified codes used for this study may be made available upon reasonable request.


\appendix

\section{Bayesian Evidence Analysis} \label{appendix}

For completeness, we compute the Bayesian evidences \citep{Trotta:2008qt} using \href{https://github.com/yabebalFantaye/MCEvidence}{MCEvidence} \citep{Heavens:2017afc}, with results listed in Table \ref{tab:bayes}. Interpreted via Jeffreys’ scale \citep{Kass01061995}, the evidences show no strong preference for interacting scenarios over their non-interacting counterparts. This is expected, as the introduction of an additional parameter ($Q$) is naturally disfavoured by model comparison metrics unless it yields a substantial improvement in fit ($\chi^2$).

However, one needs to keep in mind that this is essentially a balance between goodness of fit and model complexity, and the status keeps changing with the advent of newer data. The iDMDE models have their own physical motivations that are worth exploring, in addition to offering prospects for revisiting cosmological scenarios from a different perspective, e.g. in light of cosmological tensions, particularly with the arrival of newer data sets such as DESI DR2. Therein lies the essence of the present analysis.

\begin{table}
    \resizebox{0.5\textwidth}{!}{\renewcommand{\arraystretch}{1.4} \setlength{\tabcolsep}{15 pt}
    \begin{tabular}{c c c c c}
        \hline\hline
        \textbf{Model} & \(\Delta \log Z\) & \textbf{Interpretation} \\
        \hline
        CPL                    & 0.00  & Non-interacting (baseline) model \\
        i-CPL Phantom          & 4.74  & Strong evidence against \\
        i-CPL Non-phantom      & 4.98  & Strong evidence against \\
        JBP                    & 0.00  & Non-interacting (baseline) model \\
        i-JBP Phantom          & 5.28  & Strong evidence against \\
        i-JBP Non-phantom      & 0.76  & Weak evidence against \\
        \hline
        \hline
    \end{tabular}
    }
\caption{Model comparison using Bayesian evidences. The interpretations are based on the Jeffreys' scale.}
\label{tab:bayes}
\end{table}


\bibliographystyle{mnras}
\bibliography{mnras}

\begin{thebibliography}{}
\makeatletter
\relax
\def\mn@urlcharsother{\let\do\@makeother \do\$\do\&\do\#\do\^\do\_\do\%\do\~}
\def\mn@doi{\begingroup\mn@urlcharsother \@ifnextchar [ {\mn@doi@}
  {\mn@doi@[]}}
\def\mn@doi@[#1]#2{\def\@tempa{#1}\ifx\@tempa\@empty \href
  {http://dx.doi.org/#2} {doi:#2}\else \href {http://dx.doi.org/#2} {#1}\fi
  \endgroup}
\def\mn@eprint#1#2{\mn@eprint@#1:#2::\@nil}
\def\mn@eprint@arXiv#1{\href {http://arxiv.org/abs/#1} {{\tt arXiv:#1}}}
\def\mn@eprint@dblp#1{\href {http://dblp.uni-trier.de/rec/bibtex/#1.xml}
  {dblp:#1}}
\def\mn@eprint@#1:#2:#3:#4\@nil{\def\@tempa {#1}\def\@tempb {#2}\def\@tempc
  {#3}\ifx \@tempc \@empty \let \@tempc \@tempb \let \@tempb \@tempa \fi \ifx
  \@tempb \@empty \def\@tempb {arXiv}\fi \@ifundefined
  {mn@eprint@\@tempb}{\@tempb:\@tempc}{\expandafter \expandafter \csname
  mn@eprint@\@tempb\endcsname \expandafter{\@tempc}}}

\bibitem[\protect\citeauthoryear{Abbott et~al.}{Abbott
  et~al.}{2022}]{DES:2021wwk}
Abbott T. M.~C.,  et~al., 2022, \mn@doi [Phys. Rev. D]
  {10.1103/PhysRevD.105.023520}, 105, 023520

\bibitem[\protect\citeauthoryear{Abbott et~al.}{Abbott
  et~al.}{2023}]{Kilo-DegreeSurvey:2023gfr}
Abbott T. M.~C.,  et~al., 2023, \mn@doi [Open J. Astrophys.]
  {10.21105/astro.2305.17173}, 6, 2305.17173

\bibitem[\protect\citeauthoryear{Abbott et~al.}{Abbott
  et~al.}{2025}]{DES:2025xii}
Abbott T. M.~C.,  et~al., 2025, \mn@doi [arXiv:2503.13632]
  {10.48550/arXiv.2503.13632}

\bibitem[\protect\citeauthoryear{Abdalla et~al.}{Abdalla
  et~al.}{2022}]{Abdalla:2022yfr}
Abdalla E.,  et~al., 2022, \mn@doi [JHEAp] {10.1016/j.jheap.2022.04.002}, 34,
  49

\bibitem[\protect\citeauthoryear{Abdul~Karim et~al.}{Abdul~Karim
  et~al.}{2025}]{DESI:2025zgx}
Abdul~Karim M.,  et~al., 2025, \mn@doi [arXiv:2503.14738]
  {10.48550/arXiv.2503.14738}

\bibitem[\protect\citeauthoryear{Adame et~al.}{Adame
  et~al.}{2025}]{DESI:2024mwx}
Adame A.~G.,  et~al., 2025, \mn@doi [JCAP] {10.1088/1475-7516/2025/02/021}, 02,
  021

\bibitem[\protect\citeauthoryear{Aghanim et~al.}{Aghanim
  et~al.}{2020a}]{Planck:2019nip}
Aghanim N.,  et~al., 2020a, \mn@doi [Astron. Astrophys.]
  {10.1051/0004-6361/201936386}, 641, A5

\bibitem[\protect\citeauthoryear{Aghanim et~al.}{Aghanim
  et~al.}{2020b}]{Planck:2018vyg}
Aghanim N.,  et~al., 2020b, \mn@doi [Astron. Astrophys.]
  {10.1051/0004-6361/201833910}, 641, A6

\bibitem[\protect\citeauthoryear{Aghanim et~al.}{Aghanim
  et~al.}{2020c}]{Planck:2018lbu}
Aghanim N.,  et~al., 2020c, \mn@doi [Astron. Astrophys.]
  {10.1051/0004-6361/201833886}, 641, A8

\bibitem[\protect\citeauthoryear{Akaike}{Akaike}{1974}]{1100705}
Akaike H.,  1974, \mn@doi [IEEE Transactions on Automatic Control]
  {10.1109/TAC.1974.1100705}, 19, 716

\bibitem[\protect\citeauthoryear{Audren, Lesgourgues, Benabed  \&
  Prunet}{Audren et~al.}{2013}]{Audren:2012wb}
Audren B.,  Lesgourgues J.,  Benabed K.,   Prunet S.,  2013, \mn@doi [JCAP]
  {10.1088/1475-7516/2013/02/001}, 02, 001

\bibitem[\protect\citeauthoryear{Bansal \& Huterer}{Bansal \&
  Huterer}{2025}]{Bansal:2025ipo}
Bansal P.,  Huterer D.,  2025, \mn@doi [Phys. Rev. D] {10.1103/zypq-s6nl}, 112,
  023528

\bibitem[\protect\citeauthoryear{Bhattacharyya, Alam, Pandey, Das  \&
  Pal}{Bhattacharyya et~al.}{2019}]{Bhattacharyya:2018fwb}
Bhattacharyya A.,  Alam U.,  Pandey K.~L.,  Das S.,   Pal S.,  2019, \mn@doi
  [Astrophys. J.] {10.3847/1538-4357/ab12d6}, 876, 143

\bibitem[\protect\citeauthoryear{Blas, Lesgourgues  \& Tram}{Blas
  et~al.}{2011}]{Blas:2011rf}
Blas D.,  Lesgourgues J.,   Tram T.,  2011, \mn@doi [JCAP]
  {10.1088/1475-7516/2011/07/034}, 07, 034

\bibitem[\protect\citeauthoryear{Brinckmann \& Lesgourgues}{Brinckmann \&
  Lesgourgues}{2019}]{Brinckmann:2018cvx}
Brinckmann T.,  Lesgourgues J.,  2019, \mn@doi [Phys. Dark Univ.]
  {10.1016/j.dark.2018.100260}, 24, 100260

\bibitem[\protect\citeauthoryear{Calderon et~al.}{Calderon
  et~al.}{2024}]{DESI:2024aqx}
Calderon R.,  et~al., 2024, \mn@doi [JCAP] {10.1088/1475-7516/2024/10/048}, 10,
  048

\bibitem[\protect\citeauthoryear{Chakraborty, Ray, Das, Banerjee  \&
  Ganesan}{Chakraborty et~al.}{2024}]{Chakraborty:2024xas}
Chakraborty A.,  Ray T.,  Das S.,  Banerjee A.,   Ganesan V.,  2024, \mn@doi
  [arXiv:2403.14247] {10.48550/arXiv.2403.14247}

\bibitem[\protect\citeauthoryear{Chakraborty, Chanda, Das  \&
  Dutta}{Chakraborty et~al.}{2025}]{Chakraborty:2025syu}
Chakraborty A.,  Chanda P.~K.,  Das S.,   Dutta K.,  2025, \mn@doi
  [arXiv:2503.10806] {10.48550/arXiv.2503.10806}

\bibitem[\protect\citeauthoryear{Chevallier \& Polarski}{Chevallier \&
  Polarski}{2001}]{cpl1}
Chevallier M.,  Polarski D.,  2001, \mn@doi [Int. J. Mod. Phys. D]
  {10.1142/S0218271801000822}, 10, 213

\bibitem[\protect\citeauthoryear{Colg\'ain \& Sheikh-Jabbari}{Colg\'ain \&
  Sheikh-Jabbari}{2024}]{Colgain:2024mtg}
Colg\'ain E.~O.,  Sheikh-Jabbari M.~M.,  2024, \mn@doi [arXiv:2412.12905]
  {10.48550/arXiv.2412.12905}

\bibitem[\protect\citeauthoryear{Colg{\'a}in, Dainotti, Capozziello,
  Pourojaghi, Sheikh-Jabbari  \& Stojkovic}{Colg{\'a}in
  et~al.}{2026}]{Colgain:2024xqj}
Colg{\'a}in E.~{\'O}.,  Dainotti M.~G.,  Capozziello S.,  Pourojaghi S.,
  Sheikh-Jabbari M.~M.,   Stojkovic D.,  2026, \mn@doi [JHEAp]
  {10.1016/j.jheap.2025.100428}, 49, 100428

\bibitem[\protect\citeauthoryear{Cort\^es \& Liddle}{Cort\^es \&
  Liddle}{2024}]{Cortes:2024lgw}
Cort\^es M.,  Liddle A.~R.,  2024, \mn@doi [JCAP]
  {10.1088/1475-7516/2024/12/007}, 12, 007

\bibitem[\protect\citeauthoryear{Di~Valentino et~al.,}{Di~Valentino
  et~al.}{2021a}]{DiValentino:2021izs}
Di~Valentino E.,  et~al., 2021a, \mn@doi [Class. Quant. Grav.]
  {10.1088/1361-6382/ac086d}, 38, 153001

\bibitem[\protect\citeauthoryear{Di~Valentino et~al.}{Di~Valentino
  et~al.}{2021b}]{DiValentino:2020vvd}
Di~Valentino E.,  et~al., 2021b, \mn@doi [Astropart. Phys.]
  {10.1016/j.astropartphys.2021.102604}, 131, 102604

\bibitem[\protect\citeauthoryear{Dinda \& Maartens}{Dinda \&
  Maartens}{2025}]{Dinda:2024ktd}
Dinda B.~R.,  Maartens R.,  2025, \mn@doi [JCAP]
  {10.1088/1475-7516/2025/01/120}, 01, 120

\bibitem[\protect\citeauthoryear{Fang, Hu  \& Lewis}{Fang
  et~al.}{2008}]{Fang:2008sn}
Fang W.,  Hu W.,   Lewis A.,  2008, \mn@doi [Phys. Rev. D]
  {10.1103/PhysRevD.78.087303}, 78, 087303

\bibitem[\protect\citeauthoryear{Gavela, Hernandez, Lopez~Honorez, Mena  \&
  Rigolin}{Gavela et~al.}{2009}]{Gavela:2009cy}
Gavela M.~B.,  Hernandez D.,  Lopez~Honorez L.,  Mena O.,   Rigolin S.,  2009,
  \mn@doi [JCAP] {10.1088/1475-7516/2009/07/034}, 07, 034

\bibitem[\protect\citeauthoryear{Gavela, Lopez~Honorez, Mena  \&
  Rigolin}{Gavela et~al.}{2010}]{Gavela:2010tm}
Gavela M.~B.,  Lopez~Honorez L.,  Mena O.,   Rigolin S.,  2010, \mn@doi [JCAP]
  {10.1088/1475-7516/2010/11/044}, 11, 044

\bibitem[\protect\citeauthoryear{Gialamas, H\"utsi, Kannike, Racioppi, Raidal,
  Vasar  \& Veerm\"ae}{Gialamas et~al.}{2025}]{Gialamas:2024lyw}
Gialamas I.~D.,  H\"utsi G.,  Kannike K.,  Racioppi A.,  Raidal M.,  Vasar M.,
   Veerm\"ae H.,  2025, \mn@doi [Phys. Rev. D] {10.1103/PhysRevD.111.043540},
  111, 043540

\bibitem[\protect\citeauthoryear{Giar\`e, Najafi, Pan, Di~Valentino  \&
  Firouzjaee}{Giar\`e et~al.}{2024a}]{Giare:2024gpk}
Giar\`e W.,  Najafi M.,  Pan S.,  Di~Valentino E.,   Firouzjaee J.~T.,  2024a,
  \mn@doi [JCAP] {10.1088/1475-7516/2024/10/035}, 10, 035

\bibitem[\protect\citeauthoryear{Giar\`e, Zhai, Pan, Di~Valentino, Nunes  \&
  van~de Bruck}{Giar\`e et~al.}{2024b}]{Giare:2024ytc}
Giar\`e W.,  Zhai Y.,  Pan S.,  Di~Valentino E.,  Nunes R.~C.,   van~de Bruck
  C.,  2024b, \mn@doi [Phys. Rev. D] {10.1103/PhysRevD.110.063527}, 110, 063527

\bibitem[\protect\citeauthoryear{Giar\`e, Sabogal, Nunes  \&
  Di~Valentino}{Giar\`e et~al.}{2024c}]{Giare:2024smz}
Giar\`e W.,  Sabogal M.~A.,  Nunes R.~C.,   Di~Valentino E.,  2024c, \mn@doi
  [Phys. Rev. Lett.] {10.1103/PhysRevLett.133.251003}, 133, 251003

\bibitem[\protect\citeauthoryear{Hamana et~al.}{Hamana
  et~al.}{2020}]{Hamana:2019etx}
Hamana T.,  et~al., 2020, \mn@doi [Publ. Astron. Soc. Jap.]
  {10.1093/pasj/psz138}, 72, 16

\bibitem[\protect\citeauthoryear{Heavens, Fantaye, Mootoovaloo, Eggers,
  Hosenie, Kroon  \& Sellentin}{Heavens et~al.}{2017}]{Heavens:2017afc}
Heavens A.,  Fantaye Y.,  Mootoovaloo A.,  Eggers H.,  Hosenie Z.,  Kroon S.,
  Sellentin E.,  2017, \mn@doi [arXiv:1704.03472] {10.48550/arXiv.1704.03472}

\bibitem[\protect\citeauthoryear{Hoerning, Landim, Ponte, Rolim, Abdalla  \&
  Abdalla}{Hoerning et~al.}{2025}]{Hoerning:2023hks}
Hoerning G.~A.,  Landim R.~G.,  Ponte L.~O.,  Rolim R.~P.,  Abdalla F.~B.,
  Abdalla E.,  2025, \mn@doi [Phys. Rev. D] {10.1103/6zrh-8fmv}, 112, 023523

\bibitem[\protect\citeauthoryear{Jassal, Bagla  \& Padmanabhan}{Jassal
  et~al.}{2005}]{jbp}
Jassal H.~K.,  Bagla J.~S.,   Padmanabhan T.,  2005, \mn@doi [Phys. Rev. D]
  {10.1103/PhysRevD.72.103503}, 72, 103503

\bibitem[\protect\citeauthoryear{Kass \& Raftery}{Kass \&
  Raftery}{1995}]{Kass01061995}
Kass R.~E.,  Raftery A.~E.,  1995, \mn@doi [Journal of the American Statistical
  Association] {10.1080/01621459.1995.10476572}, 90, 773

\bibitem[\protect\citeauthoryear{Khoury, Lin  \& Trodden}{Khoury
  et~al.}{2025}]{Khoury:2025txd}
Khoury J.,  Lin M.-X.,   Trodden M.,  2025, \mn@doi [arXiv:2503.16415]
  {10.48550/arXiv.2503.16415}

\bibitem[\protect\citeauthoryear{Lesgourgues}{Lesgourgues}{2011}]{Lesgourgues:2011re}
Lesgourgues J.,  2011, \mn@doi [arXiv:1104.2932] {10.48550/arXiv.1104.2932}

\bibitem[\protect\citeauthoryear{Lewis}{Lewis}{2025}]{Lewis:2019xzd}
Lewis A.,  2025, \mn@doi [JCAP] {10.1088/1475-7516/2025/08/025}, 08, 025

\bibitem[\protect\citeauthoryear{Li et~al.}{Li et~al.}{2023}]{Li:2023azi}
Li S.-S.,  et~al., 2023, \mn@doi [Astron. Astrophys.]
  {10.1051/0004-6361/202347236}, 679, A133

\bibitem[\protect\citeauthoryear{Li, Wu, Du, Jin, Li, Zhang  \& Zhang}{Li
  et~al.}{2024}]{Li:2024qso}
Li T.-N.,  Wu P.-J.,  Du G.-H.,  Jin S.-J.,  Li H.-L.,  Zhang J.-F.,   Zhang
  X.,  2024, \mn@doi [Astrophys. J.] {10.3847/1538-4357/ad87f0}, 976, 1

\bibitem[\protect\citeauthoryear{Linder}{Linder}{2003}]{eos1}
Linder E.~V.,  2003, \mn@doi [Phys. Rev. Lett.]
  {10.1103/PhysRevLett.90.091301}, \href
  {https://ui.adsabs.harvard.edu/abs/2003PhRvL..90i1301L} {90, 091301}

\bibitem[\protect\citeauthoryear{Lodha et~al.}{Lodha
  et~al.}{2025a}]{DESI:2025fii}
Lodha K.,  et~al., 2025a, \mn@doi [arXiv:2503.14743]
  {10.48550/arXiv.2503.14743}

\bibitem[\protect\citeauthoryear{Lodha et~al.}{Lodha
  et~al.}{2025b}]{DESI:2024kob}
Lodha K.,  et~al., 2025b, \mn@doi [Phys. Rev. D] {10.1103/PhysRevD.111.023532},
  111, 023532

\bibitem[\protect\citeauthoryear{Lucca \& Hooper}{Lucca \&
  Hooper}{2020}]{Lucca:2020zjb}
Lucca M.,  Hooper D.~C.,  2020, \mn@doi [Phys. Rev. D]
  {10.1103/PhysRevD.102.123502}, 102, 123502

\bibitem[\protect\citeauthoryear{Luongo \& Muccino}{Luongo \&
  Muccino}{2024}]{Luongo:2024fww}
Luongo O.,  Muccino M.,  2024, \mn@doi [Astron. Astrophys.]
  {10.1051/0004-6361/202450512}, 690, A40

\bibitem[\protect\citeauthoryear{Mukherjee \& Sen}{Mukherjee \&
  Sen}{2024}]{Mukherjee:2024ryz}
Mukherjee P.,  Sen A.~A.,  2024, \mn@doi [Phys. Rev. D]
  {10.1103/PhysRevD.110.123502}, 110, 123502

\bibitem[\protect\citeauthoryear{Mukherjee \& Sen}{Mukherjee \&
  Sen}{2025}]{Mukherjee:2025fkf}
Mukherjee P.,  Sen A.~A.,  2025, \mn@doi [arXiv:2503.02880]
  {10.48550/arXiv.2503.02880}

\bibitem[\protect\citeauthoryear{Notari, Redi  \& Tesi}{Notari
  et~al.}{2024}]{Notari:2024rti}
Notari A.,  Redi M.,   Tesi A.,  2024, \mn@doi [JCAP]
  {10.1088/1475-7516/2024/11/025}, 11, 025

\bibitem[\protect\citeauthoryear{Pan \& Yang}{Pan \& Yang}{2023}]{Pan:2023mie}
Pan S.,  Yang W.,  2023, \mn@doi [Springer] {10.1007/978-981-99-0177-7_29}

\bibitem[\protect\citeauthoryear{Park \& Ratra}{Park \&
  Ratra}{2025}]{Chan-GyungPark:2025cri}
Park C.-G.,  Ratra B.,  2025, \mn@doi [arXiv:2501.03480]
  {10.48550/arXiv.2501.03480}

\bibitem[\protect\citeauthoryear{Park, de Cruz~P\'erez  \& Ratra}{Park
  et~al.}{2024}]{Chan-GyungPark:2024mlx}
Park C.-G.,  de Cruz~P\'erez J.,   Ratra B.,  2024, \mn@doi [Phys. Rev. D]
  {10.1103/PhysRevD.110.123533}, 110, 123533

\bibitem[\protect\citeauthoryear{Pourojaghi, Malekjani  \& Davari}{Pourojaghi
  et~al.}{2025}]{Pourojaghi:2024bxa}
Pourojaghi S.,  Malekjani M.,   Davari Z.,  2025, \mn@doi [Mon. Not. Roy.
  Astron. Soc.] {10.1093/mnras/staf037}, 537, 436

\bibitem[\protect\citeauthoryear{Roy}{Roy}{2025}]{Roy:2024kni}
Roy N.,  2025, \mn@doi [Phys. Dark Univ.] {10.1016/j.dark.2025.101912}, 48,
  101912

\bibitem[\protect\citeauthoryear{Schwarz}{Schwarz}{1978}]{10.1214/aos/1176344136}
Schwarz G.,  1978, \mn@doi [The Annals of Statistics] {10.1214/aos/1176344136},
  6, 461

\bibitem[\protect\citeauthoryear{Scolnic et~al.}{Scolnic
  et~al.}{2022}]{Scolnic:2021amr}
Scolnic D.,  et~al., 2022, \mn@doi [Astrophys. J.] {10.3847/1538-4357/ac8b7a},
  938, 113

\bibitem[\protect\citeauthoryear{Shah, Mukherjee  \& Pal}{Shah
  et~al.}{2024}]{Shah:2024rme}
Shah R.,  Mukherjee P.,   Pal S.,  2024, \mn@doi [Mon. Not. Roy. Astron. Soc.]
  {10.1093/mnras/stae2712}, 536, 2404

\bibitem[\protect\citeauthoryear{Sinha}{Sinha}{2021}]{Sinha:2021tnr}
Sinha S.,  2021, \mn@doi [Phys. Rev. D] {10.1103/PhysRevD.103.123547}, 103,
  123547

\bibitem[\protect\citeauthoryear{Sousa-Neto, Bengaly, Gonz\'alez  \&
  Alcaniz}{Sousa-Neto et~al.}{2025}]{Sousa-Neto:2025gpj}
Sousa-Neto A.,  Bengaly C.,  Gonz\'alez J.~E.,   Alcaniz J.,  2025, \mn@doi
  [arXiv:2502.10506] {10.48550/arXiv.2502.10506}

\bibitem[\protect\citeauthoryear{Torrado \& Lewis}{Torrado \&
  Lewis}{2021}]{Torrado:2020dgo}
Torrado J.,  Lewis A.,  2021, \mn@doi [JCAP] {10.1088/1475-7516/2021/05/057},
  05, 057

\bibitem[\protect\citeauthoryear{Trotta}{Trotta}{2008}]{Trotta:2008qt}
Trotta R.,  2008, \mn@doi [Contemp. Phys.] {10.1080/00107510802066753}, 49, 71

\bibitem[\protect\citeauthoryear{Wang, Abdalla, Atrio-Barandela  \&
  Pav\'on}{Wang et~al.}{2024}]{Wang:2024vmw}
Wang B.,  Abdalla E.,  Atrio-Barandela F.,   Pav\'on D.,  2024, \mn@doi [Rept.
  Prog. Phys.] {10.1088/1361-6633/ad2527}, 87, 036901

\bibitem[\protect\citeauthoryear{Wright et~al.}{Wright
  et~al.}{2025}]{Wright:2025xka}
Wright A.~H.,  et~al., 2025, \mn@doi [arXiv:2503.19441]
  {10.48550/arXiv.2503.19441}

\bibitem[\protect\citeauthoryear{Zhai, de Cesare, van~de Bruck, Di~Valentino
  \& Wilson-Ewing}{Zhai et~al.}{2025}]{Zhai:2025hfi}
Zhai Y.,  de Cesare M.,  van~de Bruck C.,  Di~Valentino E.,   Wilson-Ewing E.,
  2025, \mn@doi [arXiv:2503.15659] {10.48550/arXiv.2503.15659}

\makeatother
\end{thebibliography}

\bsp
\label{lastpage}

\end{document}